# Coupling of lattice, spin and intra-configurational excitations of $Eu^{3+}$ in $Eu_2ZnIrO_6$

Birender Singh[1#], M. Vogl[2], S. Wurmehl[2,3], S. Aswartham[2], B. Büchner[2,3] and Pradeep Kumar[1*]

*[1]School of Basic Sciences, Indian Institute of Technology Mandi, Mandi-175005, India*

*[2]Leibniz-Institute for Solid-State and Materials Research, IFW-Dresden, 01069 Dresden, Germany*

*[3]Institute of Solid-State Physics, TU Dresden, 01069 Dresden, Germany*

**Abstract:**

In $Eu_2ZnIrO_6$, effectively two atoms are active i.e. Ir is magnetically active, which results in complex magnetic ordering within the Ir sublattice at low temperature. On the other hand, although Eu is a van-vleck paramagnet, it is active in the electronic channels involving $4f^6$ crystal-field split levels. Phonons, quanta of lattice vibration, involving vibration of atoms in the unit cell, are intimately coupled with both magnetic and electronic degrees of freedom (DoF). Here, we report a comprehensive study focusing on the phonons as well as intra-configurational excitations in double-perovskite $Eu_2ZnIrO_6$. Our studies reveal strong coupling of phonons with the underlying magnetic DoF reflected in the renormalization of the phonon self-energy parameters well above the spin-solid phase ($T_N \sim 12$ K) till temperature as high as $\sim 3T_N$, evidences broken spin rotational symmetry deep into the paramagnetic phase. In particular, all the observed first-order phonon modes show softening of varying degree below $\sim 3T_N$, and low-frequency phonons become sharper, while the high-frequency phonons show broadening attributed to the additional available magnetic damping channels. We also observed a large number of high-energy modes, 39 in total, attributed to the electronic transitions between $4f$-levels of the rare-earth $Eu^{3+}$ ion and these modes shows anomalous temperature evolution as well as mixing of the crystal-field split levels attributed to the strong coupling of electronic and lattice DoF.

*#email id: birender.physics5390@gmail.com ; *email id: pkumar@iitmandi.ac.in*

## Introduction

A tremendous interest in iridates has been seen in the past few years due to a realization of the novel exotic quantum phases of matter ascribed to the strong spin-orbit coupling (SOC), electronic correlations, and their entanglement [1-12]. Double-perovskite iridium oxides of $A_2B$IrO$_6$ structure are of particular interest because of the complex quantum magnetic ground states in these materials. Their magnetic ground state is quite enigmatic, it strongly depends on the choice of $A$ and $B$-site elements, that also provide an opportunity to tune the nature of magnetic interactions. It is anticipated that such tuning of these interactions may also give rise to the much sought after quantum spin-liquid state in these double-perovskites, considered to be the holy grail of the condensed matter physics. Despite advances in an understanding of different magnetic interactions, such as Heisenberg and Kitaev-type, in these systems with the substitution of 3$d$-transition metal elements (magnetic and nonmagnetic) of varying ionic radii at $B$-site i.e. $A_2$YIrO$_6$ ($A$ = Ba, Sr) and La$_2B$IrO$_6$ ($B$ = Cu, Zn) [13-20], these iridates continue to surprise us. Where the $A$-site is occupied by an alkaline and/or rare-earth La-element; detailed investigations of complex magnetic behavior with other rare-earth elements at $A$-site are not much explored such as $Ln_2$CoIrO$_6$ ($Ln$ = Eu, Tb and Ho) and $Ln_2$ZnIrO$_6$ ($Ln$ = Sm, Nd) [21-23].

Eu$_2$ZnIrO$_6$ adopt a monoclinic double-perovskite structure with the space group $P2_1/n$ (no. 14) [24]. Eu$^{3+}$ ($4f^6$, S = 3, L = 3) is at eightfold coordinated $A$-site, whereas $B$-site is occupied by Zn$^{2+}$ ($3d^{10}$; spin S = 0), and Ir$^{4+}$ state with $5d^5$ in psudo-spin $J_{eff}$ = ½ occupies two different crystallographic sites in the corner-sharing octahedral environment. Eu$_2$ZnIrO$_6$ shows a ferromagnetic and/or canted antiferromagnetic like transition at $T_N$ ~ 12 K attributed to the ordering of spin associated with the Ir sublattice as Ir$^{4+}$ is the only magnetic ion in this compound, and Eu$^{3+}$ is a Van Vleck paramagnetic ion [24-25]. In Eu$_2$ZnIrO$_6$, Eu$^{3+}$ ion occupies $C_1$ low symmetry site, and therefore provides an opportunity to probe electronic degrees of freedom via potential electronic transitions between the different $4f$-level multiplets. In these systems, magnetic, electronic and lattice degrees of freedom attract great deal of attention as they are believed to be main controlling factors central to their underlying physics. A very important aim in the study of Ir based double-perovskite is to elucidate their ground states and the contribution of different DoF. Here, we have undertaken such a study focusing on the intricate coupling between these three DoF, i.e., lattice, magnetic and electronic. We have used inelastic light (Raman) scattering as a tool to understand the coupling between these quasi-particle excitations.

Raman spectroscopy is a sensitive and effective technique for probing the underlying lattice, magnetic and electronic degrees of freedom, and is extensively used to unravel the coupling between a range of quasiparticles in solids such as magnons, orbitons, excitons and phonons [23, 26-32]. It is basically a photon-in-photon-out process in which an incident photon of energy $\omega_i$ is inelastically scattered by the system under consideration into a photon of energy $\omega_s$. For a Stokes process, quasiparticle excitations with energy $\Delta\omega \, (Raman\,shift) = \omega_i - \omega_s$ is created in the solid. Hence, Raman scattering can be employed to investigate the crucial role of coupling between lattice and electronic degrees of freedom, involving crystal-field excitations of rare-earth elements, to the modulation of magnetic and thermodynamical properties associated with double-perovskite iridates. Within this scenario, the crystal-field structure of $4f$-levels of rare-earth allows further understanding of the fascinating ground state properties associated with these double-perovskite materials. In this paper, we report a systematic and detailed lattice-dynamics studies and intra-configurational excitations, involving crystal-field split multiplets of $Eu^{3+}$ ion for $Eu_2ZnIrO_6$ using inelastic light (Raman) scattering as a function of temperature along with the density functional theory-based calculations to understand the complex interactions between spin, lattice and electronic DoF. We observe a pronounced coupling of lattice with spin excitations and intraconfigurational electronic transitions modes of $Eu^{3+}$ ions, reflected via anomalous temperature-dependence of the self-energy parameters (i.e. phonon frequency and line-width). Phonon modes started showing softening at temperature as high as ~$3T_N$, clearly suggesting an existence of short-range magnetic correlation deep inside the paramagnetic phase. The Brillouin zone-centered calculated eigen vectors show that low-frequency phonon modes are mainly due to the vibration of Eu-atoms, while the high-frequency modes correspond to the displacement of oxygen atom associated with the $Zn/IrO_6$ octahedra. Our observation of two sharp and intense electronic transition modes from non-degenerate singlet excited state ($^5D_0$) to the singlet ground state ($^7F_0$) evident the presence of two non-equivalent $Eu^{3+}$ sites within the crystal structure. Furthermore, the assignment of symmetry of different phonon modes is done in accordance to our density functional theory-based calculations.

## Experimental and computational details

The synthesis process and detailed characterization of these high-quality polycrystalline $Eu_2ZnIrO_6$ samples are describe in detail in ref. [24]. Inelastic light (Raman) scattering measurements were performed via Raman spectrometer (LabRAM HR-Evolution) in quasi-back scattering configuration using a linearly polarized laser of 532 nm (2.33 eV) and 633 nm (1.96 eV) wavelength (energy) at low power, less than ~ 1mW, to avoid local heating effects. A 50x long working distance objective was used to focus the laser on the sample surface and the spectra was dispersed through grating with density of 600 lines/mm. A Peltier cooled charged-coupled device detector was employed to collect the dispersed light. Temperature variation was done from 4-330 K using a continuous He-flow closed-cycle cryostat (Montana instruments) with a temperature accuracy of ± 0.1 K or even better temperature accuracy.

The calculations of zone-centered phonons were done using plane-wave approach implemented in Quantum Espresso [33]. All the calculations were carried out using generalized gradient approximation with Perdew-Burke-Ernzerhof as an exchange correlation function. The plane wave cutoff energy and charge density cutoff were set to 60 and 280 Ry, respectively. Dynamical matrix and eigen vectors were determined using density functional perturbation theory [34]. The numerical integration over the Brillouin-zone was done with a 4 x 4 x 4 k-point sampling mesh in the Monkhorst-Pack grid [35]. In our calculations, we used fully relaxed ionic positions with experimental lattice parameters given in ref. [24].

## Results and Discussion

### A. Raman-scattering and lattice-dynamics calculations

$Eu_2ZnIrO_6$ crystals have double-perovskite monoclinic structure belonging to space group $P2_1/n$ (No. 14) [24]. The factor-group analysis predicts a total of 60 modes in the irreducible representation, out of which 24 are Raman active and 36 correspond to infrared active modes (see Table-I for details). Figure 1(a) shows the Raman spectra in a wide spectral range of 40-5500 $cm^{-1}$ at 4 K. The spectrum is fitted with a sum of Lorentzian functions to extract frequency ($\omega$), line-width ($\Gamma$) (Full width at half maxima - FWHM), and the integrated intensity along with the corresponding error bars of different modes (see Figs. 1(b) and 1(c)), where solid blue lines are the individual peak fits and the solid red line represent a total fit.

In order to decipher the symmetry and associated eigen vectors (atomic displacements) of different vibrational modes, we performed Brillouin zone-centered $\Gamma = (0,0,0)$ lattice-dynamics calculations using density functional theory (DFT). We note that our calculated zone-centered phonon frequencies are in very good agreement with the experimentally observed frequencies of the modes below 750 cm$^{-1}$ at 4 K, labeled as S1-S17 (see Fig. 1(b) and Table-II). Based on our first-principle calculations, we have assigned these modes below ~ 750 cm$^{-1}$, i.e. modes S1-S17, as the first-order phonon modes, while high-frequency modes in the frequency range of 1000-1400 cm$^{-1}$, i.e. S18-S20, have been assigned as second-order phonon modes. The difference between calculated and experimental phonon frequency value at 4 K is determined via absolute relative average difference formula $\left| \overline{\omega_r} \right|_{\%} = \frac{100}{N} \sum_i \left| \frac{\omega_i^{cal} - \omega_i^{exp}}{\omega_i^{exp}} \right|$, where $i$ =1, 2, − − − −, $N$ is the number of Raman active modes (17 here), and $\left| \overline{\omega_r} \right|_{\%}$ is ~ 1.5% (obtained value). Furthermore, the symmetry assignment of different phonon modes is done in accordance to our first principle lattice-dynamics calculations shown in Table-II. We notice from the visualization of phonon eigen vectors (see Fig. 2) that the low-frequency phonon modes S1-S5 arise due to the displacement of Eu-atoms, however the high-frequency phonon modes S6-S17 comprise of the bending and stretching vibrations of Zn/Ir-O bonds associated with corner-sharing Zn/IrO$_6$ octahedra. Furthermore, we also calculated full phonon dispersion and phonon density of states of Eu$_2$ZnIrO$_6$. Figure 3 (a) shows the phonon dispersion (left panel) along the $\Gamma - Z - Y - \Gamma$ high-symmetry points in the Brillouin zone, and the total phonon density of states (right panel). Figure 3 (b) illustrates the projected phonon density of states ($D(\omega)$) associated with individual atoms i.e. Eu, Zn, Ir and oxygen-atoms; and the inset shows the Brillouin zone with high-symmetry points (solid blue arrows show path of the calculated phonon dispersion). We have not observed any frequency with negative sign in the phonon dispersion that emphasizes the dynamic stability of Eu$_2$ZnIrO$_6$. The projected phonon density of states clearly shows that the density of states at low-frequency region are mainly dominated by heavy elements Eu, Zn and Ir, while the high-frequency density of states are associated with the oxygen-atoms. Apart from the first-order phonon modes, we observed a large number of additional modes at high-energy in the spectral range of 1400-4900 cm$^{-1}$, labeled as P1-P39 shown in Figure 1(c), attributed to the intraconfigurational transition modes of $4f$ levels of Eu$^{3+}$ ions, the detailed discussion is presented in sections C and D.

### B. Temperature dependence of the phonon modes

In this section, we will be focusing on understanding the temperature evolution of the first as well as second-order phonon modes. First, we will discuss the first-order phonon modes, i.e. modes below $\sim$ 750 cm$^{-1}$. As the temperature is lowered, conventionally the mode frequencies are expected to show blue-shift. At the same time, with decreasing temperature, anharmonic phonon-phonon interaction reduces which results in an increase of phonon lifetime ($\alpha$ 1/ FWHM) or a decrease of FWHM [14]. Figure 4 shows the temperature-dependence of frequency and linewidth of the first-order phonon modes. The following important observations can be made: (i) Low-frequency phonon mode S1 ($\sim$ 71 cm$^{-1}$) exhibits nearly temperature independent behavior down to $\sim$ 100 K, which gradually decreases with decrease in temperature up to $\sim$ 40 K. On further lowering the temperature, mode frequency exhibits a sharp decrease down to the lowest recorded temperature, while the corresponding damping constant shows normal behavior till $\sim$ 100 K, below which it exhibits a change of slope. (ii) All other phonon modes, S2, S4, S5, S8, S10, S11 and S17 exhibit normal temperature dependence down to $\sim$ 40 K, i.e. increase (decrease) in the mode frequency (linewidth) with decreasing temperature, and below $\sim$ 40 K these modes exhibit renormalization, i.e. anomalous phonon softening. (iii) Linewidth of the high-frequency phonon modes S8, S10, S11 and S17 exhibits line-broadening at low-temperature below $\sim$ 40 K, while minimal line-narrowing with change in slope is observed for the low-frequency phonon modes S2, S4 and S5 around this temperature. Appearance of anomaly in the phonon self-energy parameters within the spin-solid phase, i.e. long-range magnetically ordered phase, of magnetic materials is generally attributed to the entanglement of lattice with underlying spin degrees of freedom via strong spin-phonon coupling [26, 28, 31-32, 36].

Before discussing the origin of anomaly in the phonon self-energy parameters, we quantify the effect of anharmonicity on phonon modes by fitting the frequency and linewidth of the optical phonon modes with anharmonic phonon-phonon interaction model by considering the three ($\omega_1 = \omega_2 = \frac{\omega_o}{2}$) and four-phonons ($\omega_1 = \omega_2 = \omega_3 = \frac{\omega_o}{3}$) decay processes in the temperature range of 40-330 K, given as [37]:

$$\omega(T) = \omega_o + A(1 + \frac{2}{e^x - 1}) + B(1 + \frac{3}{e^y - 1} + \frac{3}{(e^y - 1)^2}) \qquad \text{......} (1)$$

and

$$\Gamma(T) = \Gamma_o + C(1 + \frac{2}{e^x - 1}) + D(1 + \frac{3}{e^y - 1} + \frac{3}{(e^y - 1)^2}) \qquad \text{......} (2)$$

where $\omega_o$ and $\Gamma_o$ are the mode frequency and linewidth at absolute zero temperature, respectively, and $x = \frac{\hbar \omega_o}{2 k_B T}$ ; $y = \frac{\hbar \omega_o}{3 k_B T}$, while A, B, C and D are constants. The parameters obtained from fitting are summarized in Table-II. We notice that the obtained values of constant parameters 'A' and 'C' are linked to three-phonon anharmonic decay processes, which dominate for all the observed phonon modes. Solid red lines in the temperature range of 40-330 K represent the anharmonic model fit to the experimental data and red solid lines below 40 K are the curve plotted using linear extrapolation method (see Fig. 4). Considering the anharmonic phonon-phonon interaction picture, the linewidth is expected to be higher with the increasing phonon energy, as also observed here (see Fig. 4). Such increase in the linewidth with increased phonon energy may be understood by the availability of more decay paths into phonons with equal and opposite wave-vector. The lowest observed phonon mode (S1) has only acoustic phonon mode channel to decay into and hence it is the sharpest, but higher energy modes will have the decay channel of the all the available lower energy phonons including the acoustic one, and hence are broader.

Phonon-phonon anharmonic model fits are in very good agreement with the observed change in the frequencies and linewidths in the temperature range of 40-330 K hinting that temperature evolution in this temperature is mainly governed by the lattice degrees of freedom. However, we noticed a pronounced deviation of the frequencies and linewidths of the phonon modes from the curve estimated by anharmonic phonon-phonon interaction model at low-temperature below ~ 40 K, indicating that this behavior can not be described within this anharmonic phonon-phonon interaction picture. For phonons in the spin-solid phase below $T_N$, an additional decay channel is expected into pairs of magnons of equal and opposite wave-vector. These observed anomalies below ~ 40 K may be understood by taking into account the interaction of phonons with the underlying magnetic degrees of freedom. The change in the phonon frequency due to interaction of lattice with spin degrees of freedom is given as [38] $\Delta \omega \approx \omega_{sp}(T) - \omega_o(T) = \lambda_{sp} < S_i.S_j >$, where $\omega_o(T)$ corresponds to the bare phonon frequency, i.e. phonon frequency without spin-phonon coupling; $\lambda_{sp} (= \frac{\partial^2 J_{ij}}{\partial u^2}$, where $J_{ij}$ is the super-exchange coupling integral) is spin-phonon coupling coefficient, which may be positive or negative, and is distinct for each phonon mode, and $< S_i.S_j >$ is the scalar spin-spin correlation function. The spin-spin correlation function is related to the order-parameter $\Phi(T)$, and within the mean field theory the $< S_i.S_j >$ is given as:

$< S_i . S_j > = - S^2 \Phi (T)$, therefore the phonon frequency deviation ($\Delta \omega$) due to spin-phonon coupling becomes [39]:

$$\Delta \omega = - \lambda_{sp} S^2 \Phi (T)$$

$$\dots \dots (3)$$

where $S$ is spin on the magnetic ion (here, we have taken $S (J_{eff}$; pseudospin$) = 1/2$, for $Ir^{4+}$ ion), and the order-parameter ($\Phi$) is given as, $\Phi (T) = 1 - \left( \dfrac{T}{T_N^*} \right)^{\gamma}$, where $\gamma$ is a critical exponent. Here, we have kept $T_N^*$ as a variable instead of keeping its value fixed i.e. ~ 12 K obtained via transport measurements because we do see significant modes renormalization well above the spin-solid phase (~ 12 K). The estimated value of spin-phonon coupling constant obtained via fitting phonon frequency deviation ($\Delta \omega$) with eq$^n$ (3) for prominent phonon modes is tabulated in Table-II. The highest obtained value of spin-phonon coupling constant is ~ 8 cm$^{-1}$. This large value of spin-phonon coupling constant indicates the strong coupling of lattice with spin degrees of freedom. However, the observation of anomalous phonon softening and linewidth broadening/narrowing well above the long-range magnetic ordering temperature ($T_N$ ~ 12K) suggests the presence of short-range spin-spin correlations deep into the paramagnetic phase, and hence broken spin rotational symmetry. We also observed that low-energy phonon modes show continuous linewidth narrowing below ~ 40 K, though with a change in slope; on the other hand, high-energy phonon modes show linewidth broadening below this temperature. This opposite behavior suggests that magnetic dispersion branches are comparable or higher in energy than these low-energy phonons, and hence these additional decay channels are mostly available only to the higher energy phonons to decay into, which showed linewidth broadening. We hope that our studies will inspire a detailed theoretical study of magnetic dispersion in these systems to quantitively understand these anomalies.

This section will discuss the origin of the high-frequency phonon modes (i.e. S18-S20) observed in the frequency range of 1000-1400 $cm^{-1}$. Considering the frequency range of these modes and the fact that these are much above the observed first-order phonon modes limiting frequency ( ~ 700 $cm^{-1}$), these modes are attributed to the second-order phonon modes corresponding to the S15-S17 first-order modes. Their origin in the intraconfigurational transitions of $Eu^{3+}$ ion is ruled out as they are also observed at the same position (i.e. Raman shift) with different excitation laser i.e. 633 nm. Second-order phonon modes are generally broader than their first-order counterparts because they involve the phonons over the entire Brillouin Zone with major contribution from the region of higher density of states. Specifically, the breadth of a second-order Raman band is governed by the dispersion of the first-order phonon modes; and also, their peak frequencies are not necessarily double of those of the first-order phonons at the gamma point. We have assigned the second-order modes S18-S20 as overtone of first-order modes S15-S17, respectively. We note that, these second-order modes may also be assigned as possible combination of the different first-order modes, e.g. S20 may be assigned as combination of S16 and S17. Figure 5 shows the temperature evolution of the peak frequency, FWHM and intensity of the prominent second-order mode i.e. S20. The following observations can be made: (i) Mode frequency and linewidth shows normal temperature-dependence i.e. increase (decrease) in frequency (FWHM) with decreasing temperature, and intensity shows an anomalous increase with decreasing temperature. The observed increase in intensity may arise because of the increased resonance effect with decreasing temperature. The change in the frequency of mode S20 ($\Delta\omega \sim 12\,cm^{-1}$) is almost double as that of first-order mode S17 ($\Delta\omega \sim 6\,cm^{-1}$) in the entire temperature range, and the linewidth at low-temperature is more than double of the linewidth of mode S17.

### C. Intraconfigurational modes of Eu$^{3+}$ in Eu$_2$ZnIrO$_6$

Apart from the first and second-order phonon modes, we observed a large number of modes in the spectral range of 1400-4900 cm$^{-1}$ (see Figs. 1(a) and 1(c)). Modes P1-P39 in the spectral range of 1400-4900 cm$^{-1}$ are attributed to the intraconfigurational electronic transitions of 4$f$-levels of Eu$^{3+}$ ions in Eu$_2$ZnIrO$_6$, where Eu$^{3+}$ ions occupy $C_1$ local site symmetry surrounded by eight oxygen atoms. Because of the $C_1$ local site symmetry, the degeneracy of the free-ion levels of Eu is completely lifted, so that (2$J$ + 1) levels are expected for each $J$ term. Figure 6 (a) shows the temperature evolution of these intraconfigurational modes in the temperature range of 4 to 330 K excited with 532 nm laser. In Eu$^{3+}$ ion, the ground state $^7F_J$ is well separated from the excited states $^5D_J$, as a result the transitions can be assigned by keeping $J$ as a good quantum number. The observed set of spectral lines in the absolute energy range of 17500-13800 cm$^{-1}$ are attributed to the transitions from first excited state ($^5D_0$) to the ground state $^7F_J$ ($J$ = 0, 1, 2, 3 and 4) multiplets of Eu$^{3+}$ ion (see Fig. 6 (a)). We have assigned the spectral band centered around (in absolute energy scale, see Figs. 6 (a)) ~ 17200 cm$^{-1}$ (P1-P3), 16800 cm$^{-1}$ (P4-P11), 16100 cm$^{-1}$ (P12-P20), 15250 cm$^{-1}$ (P21-P29) and 14150 cm$^{-1}$ (P30-P39) as transition from high-energy $^5D_0$ level to $^7F_0$, $^7F_1$, $^7F_2$, $^7F_3$ and $^7F_4$, respectively (see the schematic representation of the energy levels [40-41] of Eu$^{3+}$ ion, Fig. 6 (b)). Raman active excitations are independent of the incident photon energy, and appear at the same position in the Raman shift even with the different energy of excitation source. On the other hand, emission or absorption peak position is shifted equal to the separation between the two incident laser energies, whereas their absolute position in energy remains same. To confirm the electronic origin of these high-energy modes, we excited the spectrum using 633 nm laser (see inset of Fig.6 (a)), and only the modes below ~ 1400 cm$^{-1}$ are observed at the same position in Raman shift (not shown here), confirming the electronic transition origin of the high-energy modes (P1-P39). However, the spectrum (see inset of Fig. 6 (a)) is very weak and shows a band around ~ 14100 cm$^{-1}$ (absolute energy scale) corresponding to $^5D_0 \rightarrow ^7F_4$ transition. The observed feeble intensity of the electronic modes using 633 nm (15803 cm$^{-1}$) may be because the laser energy is much below (~ 1500 cm$^{-1}$) the first excited state $^5D_0$ (~ 17300 cm$^{-1}$) energy, and to have transition from first excited state to the ground state it needs the phonon absorption of the energy scale of ~ 1500 cm$^{-1}$.

We note that the intensity of $^5D_0 \rightarrow ^7F_J$ transition bands is remarkably quenched with decreasing temperature (see inset of Fig. 6 (a) for prominent bands). A similar decrease in the band intensity is also reported for other rare-earth based systems [41-43]. This temperature evolution of the intensity may be understood by the fact that with decreasing temperature non-radiative decay from $^5D_1$ to $^5D_0$ decreases significantly. As the incident photon energy 532 nm (18796 cm$^{-1}$) in the present case is close to the $^5D_1$ level, by absorption of the available thermal phonons via electron-phonon coupling it may easily reach $^5D_1$ level. However, with decreasing temperature, thermal population of the phonons decreases sharply, as a result the density of the $^5D_1$ state also decreases and so does the non-radiative decay rate to the $^5D_0$ state, which may finally results into decreased density of $^5D_0$ state, and hence the intensity with decreasing temperature. The transition between the electronic levels may be classified as electric-dipole (ED), magnetic-dipole (MD), and quadrupole mediated transitions; each with a different transition rate and selection rule, the strongest rate is for electric-dipole mediated one followed by magnetic-dipole and then higher order ones. Within the Judd-Ofelt theory the transition from $^5D_0 \rightarrow ^7F_2$ is ED mediated and is generally strongest [44-46]. On the other hand, $^5D_0 \rightarrow ^7F_1$ is MD mediated and is expected to be weak. However, transition from $^5D_0 \rightarrow ^7F_3$ is both ED and MD forbidden but higher order such as quadrupole as well as crystal-field mixing of the levels may contribute to these transitions, as a result these transitions are usually weak in intensity. Generally, $^5D_0 \rightarrow{}^7 F_0$ transition is not allowed except in cases where crystal-field active ion has special symmetry such as $C_s$, $C_1$ (as in the present case), $C_2$, $C_3$, $C_4$, $C_6$, $C_{2v}$, $C_{3v}$, $C_{4v}$ and $C_{6v}$, and are expected to be weak. Interestingly, our observations evidenced that most intense transition indeed occurs from $^5D_0 \rightarrow{}^7 F_2$ centered around ~ 16100 cm$^{-1}$, and transitions corresponds to the $^5D_0 \rightarrow{}^7 F_{0/1/3}$ are weak inline with the theory.

The high-energy spectrum is fitted with a sum of Lorentzian functions (see Fig. 1(c)) and extracted peak position and linewidth of the prominent modes at 4 K are tabulated in Table-III. Interestingly, the linewidth of the modes in the bands corresponding to $^5D_0 \rightarrow{}^7 F_{1/2/3}$ transitions shows continuous increase as one goes towards lower energy side, and modes corresponding to the $^5D_0 \rightarrow{}^7 F_4$ transition band show a minimal increase (see Table-III). This characteristic feature of the linewidth arises because of the transition to different levels of multiplets within the same crystal-field level, the narrow modes correspond to the transition to the lowest energy level of the multiplet and the broad modes indicate the transition to the high-energy levels of the same multiplet. This behavior of the linewidth also reflects significant coupling of lattice with electronic degrees of freedom, as the linewidth broadening is attributed to a relaxation to the available lower energy levels via emission of the phonons [47]. We observed three modes in the energy range of $^5D_0 \rightarrow{}^7 F_0$ transition at around ~ 17,300 cm$^{-1}$ (P1) and ~ 17,200 cm$^{-1}$ (P3) along with a very weak shoulder mode around ~ 17,235 cm$^{-1}$ (P2), at 4 K and the linewidth of modes are noted to be sharpest in the spectrum. The observation of more than one peak in this energy range suggests the non-equivalent crystallographic sites of Eu$^{3+}$ ion [48]. As in the case of equivalent Eu$^{3+}$ sites, only one transition is allowed due to non-degenerate nature of the excited ($^5D_0$) and ground ($^7F_0$) state multiplets. P2 and P3, separated by ~ 35 cm$^{-1}$ may be due to two equivalent sites of Eu$^{3+}$. We also observed a third peak, and it may have its origin in the transition from $^5D_1 \rightarrow^7 F_J$ levels. As $^5D_0 \rightarrow^7 F_0$ transition in the present case is allowed by the local symmetry group ($C_l$), the intensity gained by this transition band is attributed to the mixing of $^7F_0$ and $^7F_2$ crystal field levels. The estimated intensity ratio of the bands from $^5D_0 \rightarrow^7 F_0$ to $^5D_0 \rightarrow^7 F_2$ is ~ 0.012, which is quite high suggesting $J$-mixing. The intensity ratio of these transitions is also related as [49-50]:

$$\frac{I\,(^5D_0 \rightarrow^7 F_0)}{I\,(^5D_0 \rightarrow^7 F_2)} = \frac{4(B_0^2)^2}{75(\Delta_0^2)^2}$$

where, $B_0^2$ is the second-order crystal-field parameter and is associated with the environment of Eu$^{3+}$ ions and $\Delta_0^2$ corresponds to the energy difference between $^7F_0$ and $^7F_2$ states, $\Delta_0^2$ ~ 830 cm$^{-1}$ in Eu$_2$ZnIrO$_6$. Estimated value of the second-order crystal-field parameter is $B_0^2$ ~ 394 cm$^{-1}$ (~ 50 meV), which is quite high indicating strong mixing of $^7F_0$ and $^7F_2$ levels.

### D. Temperature dependence of the intraconfigurational modes of Eu$^{3+}$ ion

Now we will discuss the temperature-dependence of energy and linewidth of the prominent high-energy modes i.e. P1, P3, P4, P7, P14-P17, P19, P30, P32, P34, P36 and P37 as shown in Figure 7. Following observations can be made: (i) The energy of modes P1, P3, P7, P16, P17, P19, P36 and P37 exhibit red-shift with decreasing temperature. (ii) However, the energy of other intraconfigurational modes P4, P14, P15, P30-P32 and P34 exhibit blue-shift (shows a similar temperature-dependence as that of phonon modes) with decreasing temperature down to lowest recorded temperature 4 K. (iii) The linewidth of $^5D_0 \rightarrow^7 F_0$ transition mode P1 exhibits a change in slope around ~ 100 K with a jump of around ~ 7 cm$^{-1}$ (line-broadening) that indicates a reduced life time of transition, a similar reduction in the life time of this transition is also reported in previous studies of $Eu^{3+}$ in emission decay curves, where emission decay time changes from ∼ 2.0 ms to ∼ 0.04 µs (i.e. fast decay at low-temperature) [48]. (iv) The linewidth of all other intraconfigurational modes shows line-narrowing with decreasing temperature. Line-narrowing with decreasing temperature may occur due to the reduction of electrostatic potential at Eu-site because of reduced lattice vibrations. The surrounding ligands of the $Eu^{3+}$ ion generate a finite static potential, result into the so-called crystal-field interaction, and remove the degeneracy of the free-ion energy levels. In fact, at non-zero temperature, ions do move around their equilibrium position and renormalize the crystal-field interaction via phonons, and this interaction with phonons may explain the thermal dependence of the spectral linewidth and line position. Temperature-dependence of the energy of intra-configurational modes occurs due to a change in phonon frequency accompanying the optical excitation and dynamic modulation of the electronic levels by the phonons. The amount of change depends on the coupling strength of ions with the phonons. The energy shift of the intraconfigurational modes as a function of temperature is described using the Debye model of phonons within the weak ion-phonon coupling limit given as [47]:

$$E_i(T) = E_i(0) + \alpha_i \left( \frac{T}{\theta_D} \right)^4 \int_0^{\theta_D/T} \frac{x^3}{e^x - 1} dx \qquad \qquad \text{...... (4)}$$

where, $\alpha_i = \frac{3}{2} W_i \omega_D$, $\theta_D$ is the Debye temperature (975 K here), $\omega_D$ is the Debye phonon frequency in cm$^{-1}$ and $W_i$ is the ion-phonon coupling constant. The Debye model fit to the experimental temperature-dependence of mode energy is in good agreement, and the values obtained from the fit are summarized in Table-IV. The majority of the observed modes are shifted towards higher energy with decreasing temperature and are fit well with the eq$^n$. 4 suggesting that the associated energy levels move downward with increasing temperature. The transition modes following the temperature-dependent behavior similar to the phonons (blue-shift with decreasing temperature) unambiguously signals their phonon mediated nature. The value of ion-phonon parameter ($W_i$) is negative for these modes, however this value is positive for other modes showing red-shift with decreasing temperature (see Table-IV). The red-shift of some of the observed modes with decreasing temperature may be understood by so-called pushing effect or fast lowering of the terminal levels [51]. A similar red-shift has also been reported for other rare-earth elements [23, 51-53]. It has been associated with the fact that terminal levels lower faster than other levels with increasing temperature, therefore transition from higher level to these terminal levels of any set of multiplets will result in the red-shift. More theoretical work is required to understand these red-shifts and their coupling with other quasi-particles.

**Summary and conclusions**

To summarize, we report the detailed lattice-dynamics study of double-perovskite $Eu_2ZnIrO_6$ via Raman scattering as a function of temperature and density functional theory-based calculations. We find significant phonon softening and anomalous linewidth narrowing/broadening of the phonon modes well above the spin-solid phase ($T_N \sim 12$ K) up to $\sim 40$ K. Our study foretells that quantum magnetic ground state of $5d$ iridium based double-perovskite materials is the result of intricate coupling of lattice, spin and electronic degrees of freedom. Suggesting that these degrees of freedom should be treated at par with each other to write the ground state Hamiltonian for their understanding. Estimated value of the spin-phonon coupling constant ($\lambda_{sp}$) is found to be ranging from $\sim 5$ to 8 cm$^{-1}$. The high value of second-order crystal-field parameter, $B_0^2 \sim 50\ meV$, suggests strong $J$-mixing of the crystal-field split levels of $Eu^{3+}$ ion. The density functional theory based calculated zone-centered phonon mode frequencies are observed to be in very good agreement with the experimentally observed values. Our lattice dynamics studies reveal the rich physics associated with optical phonons and their coupling to electronic and magnetic degrees of freedom in this system. We believe that our studies will motivate further theoretical work for the estimation of all possible intraconfigurational transitions of $4f$-levels of $Eu^{3+}$ in $C_1$ low symmetry configuration as well as quantitatively understand the coupling between these different degrees of freedom and their contribution to the ground state.

**Acknowledgment**

PK thanks the Department of Science and Technology, India, for the grant and IIT Mandi for the experimental facilities. The authors at Dresden thank Deutsche Forschungsgemeinschaft (DFG) for financial support via Grant No. DFG AS 523/4-1 (S.A.) and via project B01 of SFB 1143 (project-id 247310070).

**References:**

[1]     B. J. Kim, Hosub Jin, S. J. Moon, J.-Y. Kim, B.-G. Park, C. S. Leem, Jaejun Yu, T. W. Noh, C. Kim, S.-J. Oh, J.-H. Park, V. Durairaj, G. Cao, and E. Rotenberg, Phys. Rev. Lett. **101**, 076402 (2008).

[2]     B. J. Kim, H. Ohsumi, T. Komesu, S. Sakai, T. Morita, H. Takagi, and T. Arima, Science **323**, 1329 (2009).

[3]     H. Watanabe, T. Shirakawa, and S. Yunoki, Phys. Rev. Lett. **105**, 216410 (2010).

[4]     Y. Yamaji, Y. Nomura, M. Kurita, R. Arita, and M. Imada, Phys. Rev. Lett. **113**, 107201 (2014).

[5]    C. Dhital, T. Hogan, W. Zhou, X. Chen, Z. Ren, M. Pokharel, Y. Okada, M. Heine, W. Tian, Z. Yamani, C. Opeil, J. S. Helton, J. W. Lynn, Z. Wang, V. Madhavan, and S. D. Wilson, Nat. Commun. **5**, 3377 (2014).

[6]    G. Cao, T. F. Qi, L. Li, J. Terzic, S. J. Yuan, L. E. DeLong, G. Murthy, and R. K. Kaul, Phys. Rev. Lett. **112**, 056402 (2014).

[7]    H. Gretarsson, N. H. Sung, M. Höppner, B. J. Kim, B. Keimer, and M. Le Tacon, Phys. Rev. Lett. **116**, 136401 (2016).

[8]    Y. Cao, Q. Wang, J. A. Waugh, T. J. Reber, H. Li, X. Zhou, S. Parham, S.-R. Park, N. C. Plumb, E. Rotenberg, A. Bostwick, J. D. Denlinger, T. Qi, Michael A. Hermele, G. Cao, and D. S. Dessau, Nat. Commun. **7**, 11367 (2016).

[9]    S. Kanungo, K. Mogare, B. Yan, M. Reehuis, A. Hoser, C. Felser, and M. Jansen, Phys. Rev. B **93**, 245148 (2016).

[10]   M. Nakayama, T. Kondo, Z. Tian, J. J. Ishikawa, M. Halim, C. Bareille, W. Malaeb, K. Kuroda, T. Tomita, S. Ideta, K. Tanaka, M. Matsunami, S. Kimura, N. Inami, K. Ono, H. Kumigashira, L. Balents, S. Nakatsuji, and S. Shin, Phys. Rev. Lett. **117**, 056403 (2016).

[11]   L. T. Corredor, G. A.-Cansever, M. Sturza, K. Manna, A. Maljuk, S. Gass, T. Dey, A. U. B. Wolter, O. Kataeva, A. Zimmermann, M. Geyer, C. G. F. Blum, S. Wurmehl, and B. Büchner, Phys. Rev. B **95**, 064418 (2017).

[12]   A. A. Aczel, J. P. Clancy, Q. Chen, H. D. Zhou, D. Reig-i-Plessis, G. J. MacDougall, J. P. C. Ruff, M. H. Upton, Z. Islam, T. J. Williams, S. Calder, and J.-Q. Yan, Phys. Rev. B **99**, 134417 (2019).

[13]   T. Dey, A. Maljuk, D. V. Efremov, O. Kataeva, S. Gass, C. G. F. Blum, F. Steckel, D. Gruner, T. Ritschel, A. U. B. Wolter, J. Geck, C. Hess, K. Koepernik, J. van den Brink, S. Wurmehl, and B. Büchner, Phys. Rev. B **93**, 014434 (2016).

[14]   B. Singh, G. A. Cansever, T. Dey, A. Maljuk, S. Wurmehl, B Büchner, and Pradeep Kumar, J. Phys.: Condens. Matter **31**, 065603 (2019).

[15]   Q. Chen, C. Svoboda, Q. Zheng, B. C. Sales, D. G. Mandrus, H. D. Zhou, J.-S. Zhou, D. McComb, M. Randeria, N. Trivedi, and J.-Q. Yan, Phys. Rev. B **96**, 144423 (2017).

[16]   S. Fuchs, T. Dey, G. A.-Cansever, A. Maljuk, S. Wurmehl, B. Büchner, and V. Kataev, Phys. Rev. Lett. **120**, 237204 (2018).

[17]   G. Cao, A. Subedi, S. Calder, J.-Q. Yan, J. Yi, Z. Gai, L. Poudel, D. J. Singh, M. D. Lumsden, A. D. Christianson, B. C. Sales, and D. Mandrus, Phys. Rev. B **87**, 155136 (2013).

[18]   K. Manna, R. Sarkar, S. Fuchs, Y. A. Onykiienko, A. K. Bera, G. A. Cansever, S. Kamusella, A. Maljuk, C. G. F. Blum, L. T. Corredor, A. U. B. Wolter, S. M. Yusuf, M. Frontzek, L. Keller, M. Iakovleva, E. Vavilova, H.-J. Grafe, V. Kataev, H.-H. Klauss, D. S. Inosov, S. Wurmehl, and B. Büchner, Phys. Rev. B **94**, 144437 (2016).

[19]   B. Singh, D. Kumar, K. Manna, A. K. Bera, G. A. Cansever, A. Maljuk, S. Wurmehl, B. Büchner, and P. Kumar, J. Phys.: Condens. Matter **31**, 485803 (2019).

[20]   A. A. Aczel, A. M. Cook, T. J. Williams, S. Calder, A. D. Christianson, G.-X. Cao, D. Mandrus, Y.-B. Kim, and A. Paramekanti, Phys. Rev. B **93**, 214426 (2016).

[21]   X. Ding, B. Gao, E. Krenkel, C. Dawson, J. C. Eckert, S.-W. Cheong, and V. Zapf, Phys. Rev. B **99**, 014438 (2019).

[22]   B. Singh, M. Vogl, S. Wurmehl, S. Aswartham, B. Büchner, and P. Kumar, Phys. Rev. Research **2**, 013040 (2020).

[23]   B. Singh, M. Vogl, S. Wurmehl, S. Aswartham, B. Büchner, and P. Kumar, Phys. Rev. Research **2**, 023162 (2020).

[24]   M. Vogl, R. Morrow, A. A. Aczel, R. B. Rodriguez, A. U. B. Wolter, S. Wurmehl, S. Aswartham, and B. Büchner, Phys. Rev. Materials **4**, 054413 (2020).

[25]   Y. Takikawa, S. Ebisu, and S. Nagata, J. Phys. Chem. Solids **71**, 1592 (2010).

[26]   P. Kumar, A. Bera, D. V. S. Muthu, S. N. Shirodkar, R. Saha, A. Shireen, A. Sundaresan, U. V. Waghmare, A. K. Sood, and C. N. R. Rao, Phys. Rev. B **85**, 134449 (2012).

[27]   P. Kumar, D. V. S. Muthu, L. Harnagea, S. Wurmehl, B. Büchner, and A K Sood, J. Phys.: Condens. Matter **26**, 305403 (2014).

[28]   M. G. Cottam and D. J. Lockwood, Light Scattering in Magnetic Solids, Wiley, New York (1986).

[29]   M. Cardona, Light scattering in solids III Topics in Applied Physics, ed M Cardona and G Guntherdot (Berlin: Springer), 1982.

[30]   P. Kumar, K. P. Ramesh, and D. V. S. Muthu, AIP Advances **5**, 037135 (2015).

[31]   B. Singh, S. Kumar, and P. Kumar, J. Phys.: Condens. Matter **31**, 395701 (2019).

[32]   P. Kumar, A. Kumar, S. Saha, D. V. S. Muthu, J. Prakash, S. Patnaik, U. V. Waghmare, A. K. Ganguli, and A. K. Sood, Solid State Commun. **150**, 557 (2010).

[33]   P. Giannozzi, S. Baroni, N. Bonini, M. Calandra, R. Car, C. Cavazzoni, D. Ceresoli, G. L. Chiarotti, M. Cococcioni et al., J. Phys.: Condens. Matter **21**, 395502 (2009).

[34]   J. P. Perdew, A. Ruzsinszky, G. I. Csonka, O. A. Vydrov, G. E. Scuseria, L. A. Constantin, X. Zhou, and K. Burke, Phys. Rev. Lett. **100**, 136406 (2008).

[35]   P. Giannozzi, S. de Gironcoli, P. Pavone, and S. Baroni, Phys. Rev. B **43**, 7231 (1991).

[36]   M. G. Cottam and D. J. Lockwood, Low Temp. Phys. **45**, 78 (2019).

[37]   M. Balkanski, R. F. Wallis, and E. Haro, Phys. Rev. B **28**, 1928 (1983).

[38]   D. J. Lockwood and M. G. Cottam, J. Appl. Phys. **64**, 5876 (1988).

[39]   C.-H. Hung, P.-H. Shih, F.-Y. Wu, W.-H. Li, S. Y. Wu, T. S. Chan, and H.-S. Sheu, J. Nanosci. Nanotechnol. **10**, 4596 (2010).

[40]   S. Taboada, A. de Andrés, and R. Sáez-Puche, J. Alloys Compd. **275**, 279 (1998).

[41]   S. Taboada, A. de Andrés, and J. E. Munoz-Santiuste, J. Phys.: Condens. Matter. **10**, 8983 (1998).

[42]   S. Taboada, J. E. Munoz-Santiuste, and A. de Andrés, J. Lumin. **72**, 273 (1997).

[43]   G. Blasse, Solid State Luminescence: Theory, Materials and Devices 1993 edited by A. H. Kitai (London: Chapman and Hall)

[44]   B. R. Judd, Phys. Rev. **127**, 750 (1962).

[45]   G.S. Ofelt, J. Chem. Phys. **37**, 511 (1962).

[46]   X. Y. Chen and G.K. Liu, J. solid state chemistry **178**, 419 (2005).

[47]   S. Hüfner, Optical Spectra of Transparent Rare Earth Compounds (Academic Press, New York, 1978)

[48]   Y. Huang, H. Lin, and H. Jin Seo, J. Electrochem. Soc. **157**, J405-J409 (2010).

[49]   S. Ray, P. Pramanik, A. Singha, and A. Roya, J. Appl. Phys. **97**, 094312 (2005).

[50]   A. K. Parchur and R. S. Ningthoujam, RSC Advances, **2**, 10859 (2012).

[51]   T. Kyshida, Phys. Rev. **185**, 500 (1969).

[52]   S. Saha, S. Prusty, S. Singh, R. Suryanarayanan, A. Revcolevschi, and A. K. Sood, J. Phys.: Condens. Matter **23**, 445402 (2011).

[53]   M. G. Beghi, C. E. Bottani, and V. Russo J. Appl. Phys. **87**, 1769 (2000).

**Table-I:** Wyckoff positions and irreducible representations of the phonon modes, at the gamma point of the Brillouin zone, of the monoclinic ($P2_1/n$; space group No. 14) double-perovskite Eu$_2$ZnIrO$_6$. $\Gamma_{total}$, $\Gamma_{Raman}$ and $\Gamma_{\inf rared}$ represent the total normal modes, the Raman and the infrared active modes, respectively.

| $\boldsymbol{P2_1/n}$; space group No. 14 | | |
|---|---|---|
| Atom | Wyckoff site | Mode decomposition |
| Eu | 4e | $3A_g + 3A_u + 3B_g + 3B_u$ |
| Zn | 2d | $3A_u + 3B_u$ |
| Ir | 2c | $3A_u + 3B_u$ |
| O(1) | 4e | $3A_g + 3A_u + 3B_g + 3B_u$ |
| O(2) | 4e | $3A_g + 3A_u + 3B_g + 3B_u$ |
| O(3) | 4e | $3A_g + 3A_u + 3B_g + 3B_u$ |
| | $\Gamma_{total} = 12A_g + 18A_u + 18B_u + 12B_g$ | |
| | $\Gamma_{Raman} = 12A_g + 12B_g$ and $\Gamma_{\inf rared} = 18A_u + 18B_u$ | |

**Table-II:** List of the experimentally observed Raman active phonon mode frequencies at 4 K (monoclinic space group $P2_1/n$) with 532 nm excitation laser along with the DFT based calculated zone centered frequencies. Fitting parameters obtained using equations as described in the text, and estimated values of spin-phonon coupling constant ($\lambda_{sp}$) for the prominent phonon modes. The symmetry assignment of phonon modes is done in accordance to our lattice-dynamics calculations. The low-frequency modes S1-S5 are mainly attributed to the vibrations of Eu-atom, while high-frequency modes S6-S17 are associated with the bending and stretching of Zn/Ir-O bonds in the Zn/IrO$_6$ octahedral units. The calculated frequencies not observed experimentally are as 93.5 (B$_g$), 207.4 (B$_g$), 208.1 (A$_g$), 293.0 (A$_g$), 356.2 (A$_g$), 411.7 (B$_g$) and 521.0 (A$_g$). Units are in cm$^{-1}$.

| Mode Assignment | Exp. ω (4 K) | DFT ω | $\lambda_{sp}$ | Fitted Parameters | | | | | |
|---|---|---|---|---|---|---|---|---|---|
| | | | | $\omega_0$ | A | B | $\Gamma_0$ | C | D |
| S1-B$_g$ (Eu) | 71.1 | 87.8 | | | | | | | |
| S2-A$_g$ (Eu) | 101.6 | 95.9 | 8.2 | 103.8 ± 0.2 | -0.19 ± 0.07 | -0.003 ± 0.002 | 9.2 ± 0.7 | 0.39 ± 0.03 | 0.005 ± 0.001 |
| S3-A$_g$ (Eu) | 125.6 | 119.8 | | 126.6 ± 0.2 | 0.08 ± 0.01 | -0.006 ± 0.002 | 5.1 ± 0.4 | -0.04 ± 0.02 | 0.003 ± 0.001 |
| S4-A$_g$ (Eu) | 136.6 | 125.7 | 5.9 | 137.6 ± 0.2 | 0.27 ± 0.12 | -0.016 ± 0.003 | 4.2 ± 0.3 | 0.21 ± 0.01 | 0.008 ± 0.004 |
| S5-B$_g$ (Eu) | 160.8 | 152.1 | 7.4 | 162.8 ± 0.2 | -0.18 ± 0.14 | -0.017 ± 0.005 | 4.3 ± 0.3 | 0.19 ± 0.02 | 0.011 ± 0.007 |
| S6-A$_g$ (O) | 237.4 | 237.9 | | 242.1 ± 1.2 | 5.32 ± 0.96 | -0.42 ± 0.05 | 20.9 ± 2.1 | 4.79 ± 1.08 | 0.187 ± 0.021 |
| S7-B$_g$ (O) | 273.7 | 284.3 | | 277.1 ± 2.6 | 5.42 ± 2.09 | -0.51 ± 0.11 | 17.4 ± 3.4 | 1.37 ± 0.49 | -0.173 ± 0.091 |
| S8-B$_g$ (O) | 306.5 | 302.1 | 7.3 | 308.3 ± 0.5 | 0.30 ± 0.11 | -0.18 ± 0.032 | 7.5 ± 1.8 | 1.72 ± 1.12 | 0.109 ± 0.015 |
| S9-B$_g$ (O) | 373.6 | 364.8 | | 375.5 ± 1.5 | 0.79 ± 1.41 | -0.37 ± 0.11 | 15.9 ± 3.3 | 1.19 ± 1.01 | 0.008 ± 0.003 |
| S10-A$_g$ (O) | 407.6 | 408.0 | 6.3 | 409.1 ± 1.1 | -0.27 ± 0.11 | -0.24 ± 0.10 | 10.1 ± 2.8 | 8.77 ± 3.12 | 0.101 ± 0.013 |
| S11-A$_g$ (O) | 425.2 | 419.7 | 5.2 | 426.1 ± 0.8 | 0.71 ± 0.21 | -0.26 ± 0.06 | 5.2 ± 2.1 | 6.73 ± 2.07 | -0.028 ± 0.017 |
| S12-B$_g$ (O) | 473.2 | 467.7 | | 473.3 ± 3.7 | 0.60 ± 0.17 | -0.40 ± 0.16 | 9.5 ± 3.1 | 5.33 ± 3.05 | 0.075 ± 0.013 |
| S13-A$_g$ (O) | 509.6 | 501.1 | | 511.5 ± 4.8 | 1.16 ± 0.51 | -0.59 ± 0.13 | 4.3 ± 1.8 | 2.94 ± 1.37 | -0.071 ± 0.007 |
| S14-B$_g$ (O) | 527.4 | 523.1 | | 532.3 ± 1.8 | -3.37 ± 1.94 | -0.26 ± 0.21 | 10.6 ± 3.3 | 7.47 ± 2.06 | 0.172 ± 0.095 |
| S15-B$_g$ (O) | 568.9 | 534.8 | | | | | | | |
| S16-B$_g$ (O) | 659.5 | 601.1 | | 653.9 ± 5.4 | 10.89 ± 4.53 | -2.52 ± 1.25 | 18.5 ± 5.1 | 10.06 ± 3.42 | -0.536 ± 0.241 |
| S17-A$_g$ (O) | 677.8 | 651.0 | 7.2 | 686.8 ± 4.4 | -6.95 ± 3.21 | -0.43 ± 0.16 | 5.4 ± 2.7 | 13.06 ± 4.13 | -0.667 ± 0.148 |
| S18 (Second-order) | 1133.6 | | | | | | | | |
| S19 (Second-order) | 1215.9 | | | | | | | | |
| S20 (Second-order) | 1321.8 | | | | | | | | |

**Table-III:** List of the frequency and linewidth of the prominent intraconfigurational transition modes of $Eu^{3+}$ in $Eu_2ZnIrO_6$, at 4 K. Units are in $cm^{-1}$.

| $^5D_0 \rightarrow {}^7F_0$ | | | $^5D_0 \rightarrow {}^7F_1$ | | | $^5D_0 \rightarrow {}^7F_2$ | | | $^5D_0 \rightarrow {}^7F_3$ | | | $^5D_0 \rightarrow {}^7F_4$ | | |
|------|------------------|------|------|------------------|------|------|------------------|-------|------|------------------|------|------|------------------|------|
| **Mode** | $\omega_{abs.}$ | $\Gamma$ | **Mode** | $\omega_{abs.}$ | $\Gamma$ | **Mode** | $\omega_{abs.}$ | $\Gamma$ | **Mode** | $\omega_{abs.}$ | $\Gamma$ | **Mode** | $\omega_{abs.}$ | $\Gamma$ |
| **P1** | 17,296 | 18.4 | **P7** | 16,933 | 39.0 | **P13** | 16,369 | 43.6 | **P24** | 15,288 | 22.9 | **P31** | 14,554 | 22.3 |
| **P3** | 17,198 | 12.9 | **P8** | 16,868 | 53.9 | **P14** | 16,298 | 37.2 | **P25** | 15,267 | 23.0 | **P32** | 14,403 | 23.7 |
| | | | **P9** | 16,833 | 65.7 | **P15** | 16,266 | 36.6 | **P26** | 15,234 | 54.0 | **P34** | 14,245 | 30.2 |
| | | | **P10** | 16,787 | 80.7 | **P16** | 16,210 | 79.6 | | | | **P36** | 14,146 | 24.5 |
| | | | | | | **P17** | 16,057 | 84.1 | | | | **P37** | 14,111 | 23.9 |
| | | | | | | **P18** | 15,965 | 112.4 | | | | | | |

**Table-IV:** List of the parameters extracted from fitting the temperature dependence of intraconfigurational mode energies of $Eu^{3+}$ in $Eu_2ZnIrO_6$ as described in the text. The value of $\theta_D = 975$ K.

| Mode | $\omega(0)$ | $\alpha_i$ | $W_i$ | Mode | $\omega(0)$ | $\alpha_i$ | $W_i$ |
|------|-------------|-----------|-------|------|-------------|-----------|-------|
| **P1** | 17,296 | +82.6 | +0.08 | **P19** | 15,889 | +117.7 | +0.11 |
| **P3** | 17,198 | +232.6 | +0.23 | **P30** | 14,609 | -223.9 | -0.22 |
| **P4** | 17,127 | -54.6 | -0.05 | **P31** | 14,555 | -126.5 | -0.12 |
| **P7** | 16,932 | +29.9 | +0.03 | **P32** | 14,403 | -120.7 | -0.11 |
| **P14** | 16,297 | -29.5 | -0.03 | **P34** | 14,245 | -37.4 | -0.04 |
| **P15** | 16,266 | -22.5 | -0.02 | **P36** | 14,145 | +26.9 | +0.03 |
| **P16** | 16,209 | +57.4 | +0.06 | **P37** | 14,109 | +59.8 | +0.06 |
| **P17** | 16,057 | +85.8 | +0.06 | | | | |

**FIGURE CAPTION:**

**FIGURE 1:** (Color online) (a) Raman spectra of $Eu_2ZnIrO_6$ recorded at 4 K in the spectral range of 40-5500 $cm^{-1}$ using 532 nm laser. Yellow, light yellow and light blue shaded area indicates first-order, second-order phonon modes and intraconfigurational modes of $Eu^{3+}$ ions in $Eu_2ZnIrO_6$, respectively. (b) Raman spectra of first-order phonon modes in the spectral range of 40-750 $cm^{-1}$ fitted with a sum of Lorentzian functions, where solid thick red line indicates a total fit to the experimental data and solid thin blue line corresponds to individual peak fit; inset shows temperature evolution of the spectrum in the temperature range of 4-330 K. (c) High-energy Raman spectrum fitted with a sum of Lorentzian functions in the spectral range of 1000-4900 $cm^{-1}$. Inset shows the amplified spectrum in the frequency range of 1000-2250 $cm^{-1}$ and 3300-3800 $cm^{-1}$.

**FIGURE 2:** (Color online) Schematic representation of the eigen vectors of the first-order phonon modes S1 to S17 of $Eu_2ZnIrO_6$, where pink, blue, dark yellow and red spheres represent the Eu, Zn, Ir and Oxygen-atoms, respectively. Black arrows on the atoms indicate the direction of atomic displacement, and the magnitude of the arrow illustrates the extent of atomic vibration. a, b and c indicate the crystallographic axis.

**FIGURE 3:** (Color online) (a) (left panel) Phonon dispersion, red and blue lines represent dispersion of the Raman active $A_g$ and $B_g$ modes, respectively, broken black lines illustrate the infrared modes. (Right panel) Total phonon density of states of $Eu_2ZnIrO_6$. (b) Projected phonon density of states associated with Eu, Zn, Ir and Oxygen-atoms. Inset shows first Brillouin zone along with high symmetry path, solid blue lines indicate the path of calculated phonon dispersion.

**FIGURE 4:** (Color online) Temperature dependence of the frequency and linewidth of prominent first-order phonon modes of $Eu_2ZnIrO_6$. Solid red lines in the temperature range of 40-330 K are the fitted curves as described in the text and red lines below 40 K are the extrapolated curves. Solid yellow lines are guide to the eye and the shaded part depicts the magnetically ordered phase.

**FIGURE 5:** (Color online) Temperature dependence of frequency, linewidth and intensity of prominent second-order phonon (S20) mode of $Eu_2ZnIrO_6$.

**FIGURE 6:** (Color online) (a) Temperature evolution of the high-energy spectra of intraconfigurational modes of $Eu^{3+}$ ions in the absolute energy range of 18500-13500 $cm^{-1}$. Inset shows the spectra excited with 633 nm laser, and the temperature dependence of intensity of $^5D_0 \rightarrow\, ^7F_2$ and $^5D_0 \rightarrow\, ^7F_4$ transitions bands (solid black line is guide to the eye). (b) The schematic representation of the crystal-field splitting of energy levels of $Eu^{3+}$ ion. Vertical blue lines indicate the observed transitions.

**FIGURE 7:** (Color online) Temperature dependence of energy and linewidth of the prominent high-energy intraconfigurational modes P1, P3, P4, P7, P14-P17, P19, P30-P32, P34, P36 and P37 in the temperature range of 4-330 K. Solid red lines are the fitted curves as described in the text.

**Figure 1:**

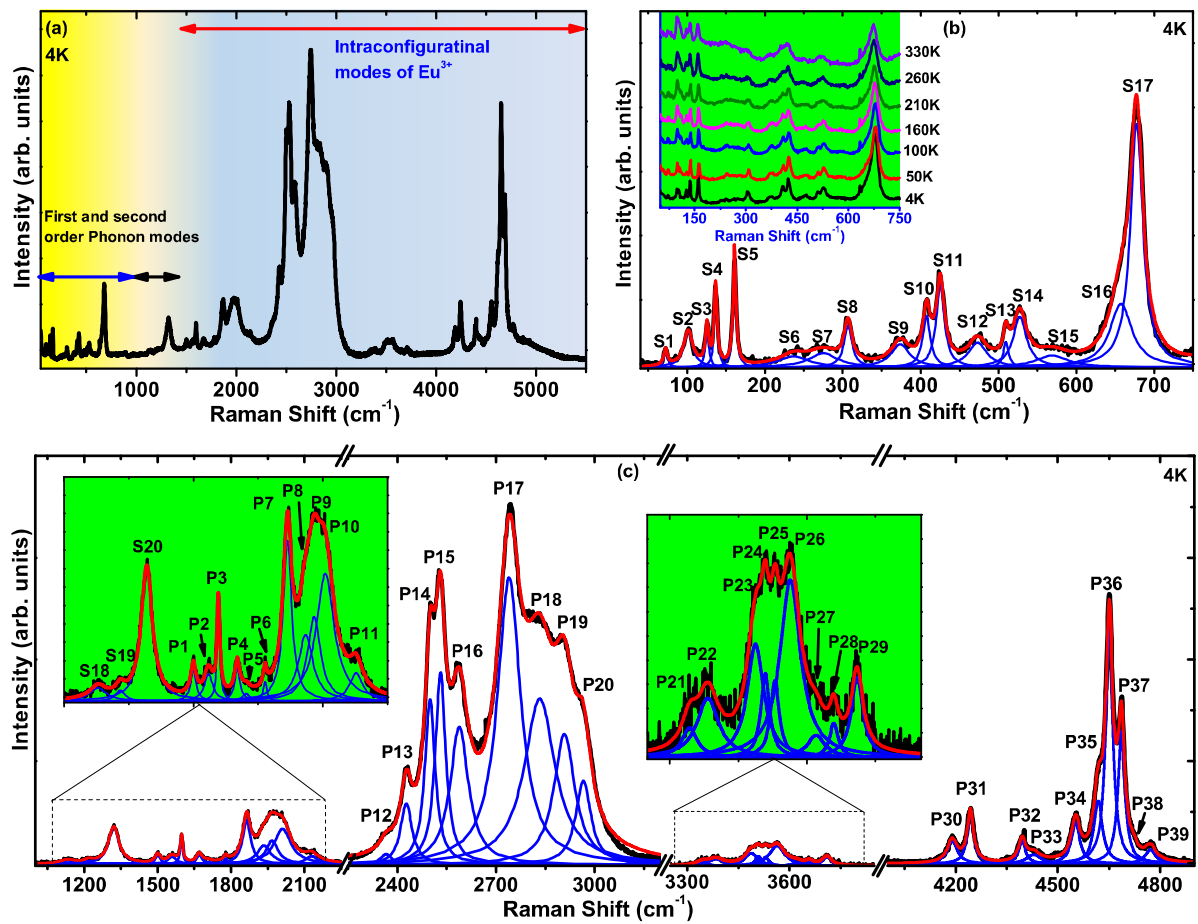

**Figure 2:**

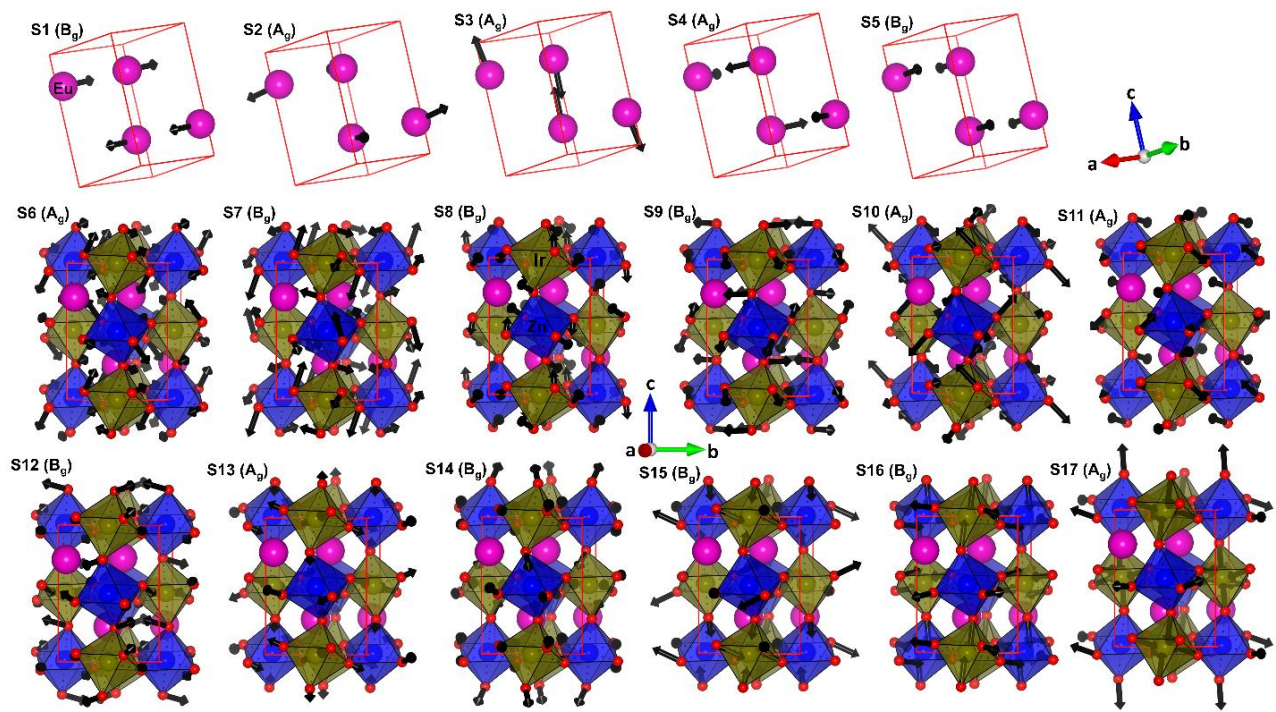

**Figure 3:**

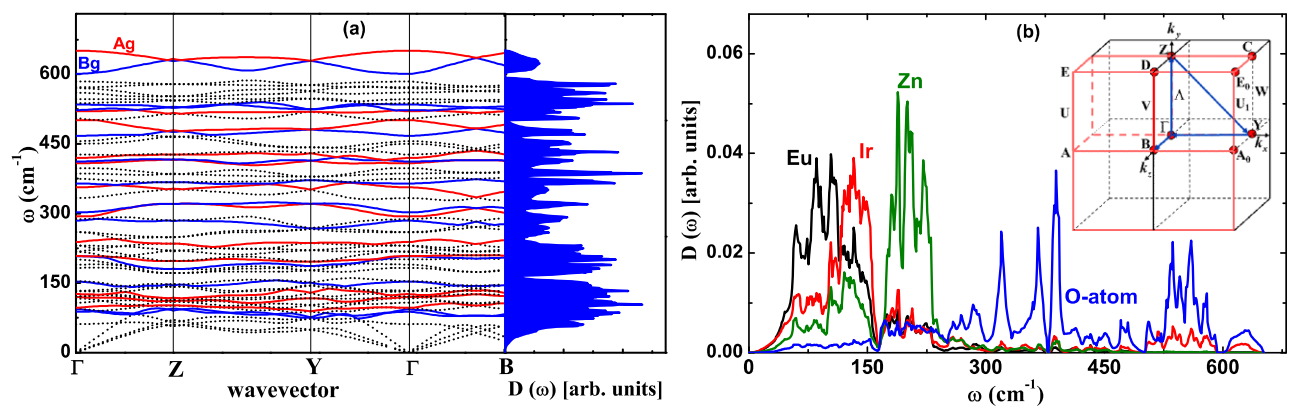

**Figure 4:**

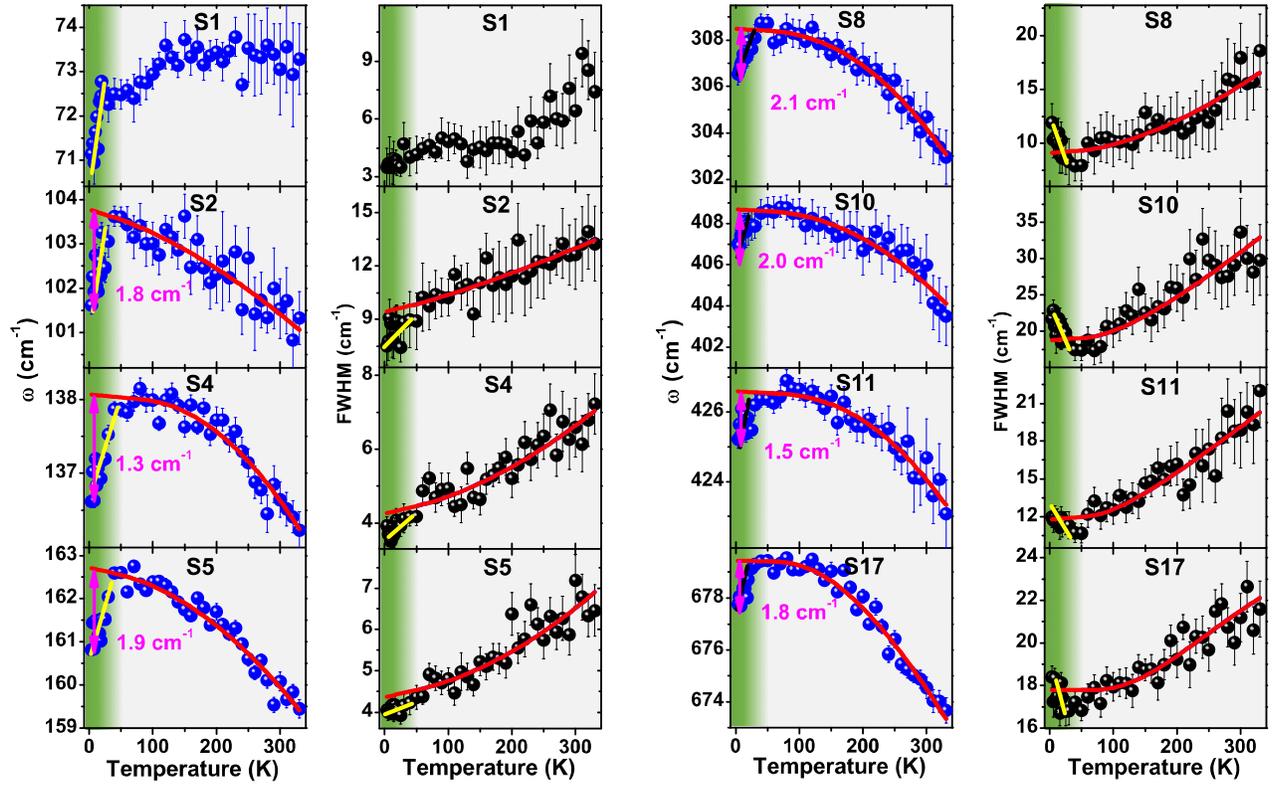

**Figure 5:**

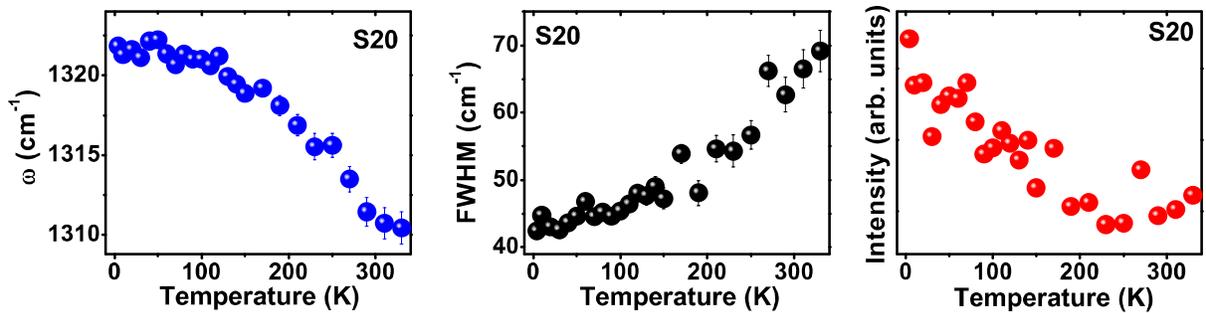

**Figure 6:**

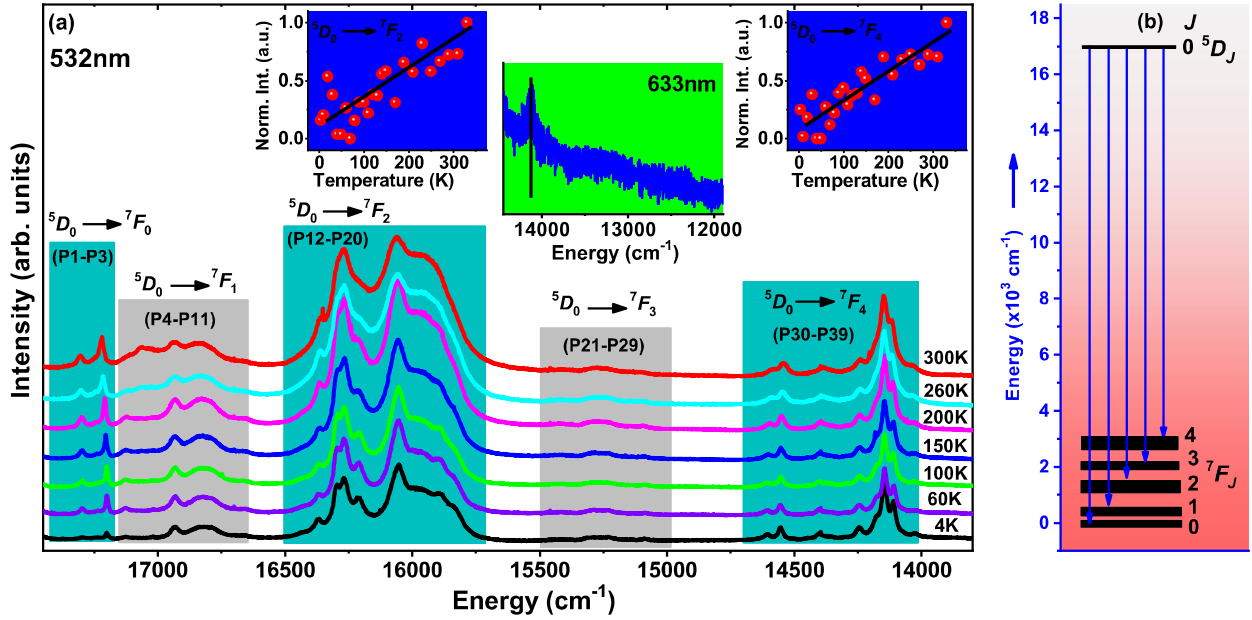

**Figure 7:**

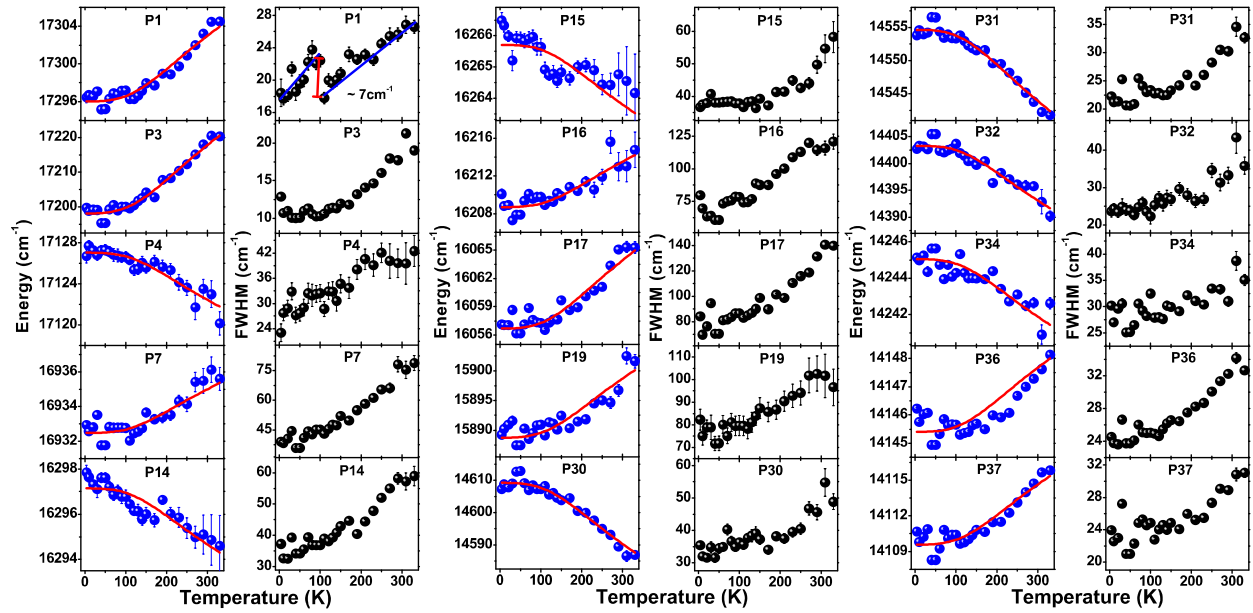